\documentclass[conference, 10pt]{IEEEtran}
\usepackage{cite}

\ifCLASSINFOpdf
\else
\fi
\usepackage{graphicx}
\usepackage{color}
\usepackage{placeins}
\usepackage{float}
\usepackage{hyperref}
\usepackage{tabularx,colortbl}
\usepackage{amsmath}

\begin{document}
%
\title{\textcolor{black}{Rigorous Analytical Model for Metasurface Microscopic Design with Interlayer Coupling}}

\author{\IEEEauthorblockN{Shahar Levy, Yaniv Kerzhner and Ariel Epstein\IEEEauthorrefmark{1},}
	\IEEEauthorblockA{Andrew and Erna Viterbi Faculty of Electrical Engineering, Technion-Israel Institute of Technology, Haifa 3200003, Israel\\
		Email: \IEEEauthorrefmark{1}epsteina@ee.technion.ac.il}}


%


\maketitle

\begin{abstract}
We present a semianalytical method for designing meta-atoms in multilayered metasurfaces (MSs), relying on a rigorous model developed for multielement metagratings. Notably, this model properly accounts for near-field coupling effects, allowing reliable design even for extremely small interlayer spacings, verified via commercial solvers. This technique forms an appealing alternative to the common full-wave optimization employed for MS microscopic design to date.
\end{abstract}


%
\IEEEpeerreviewmaketitle

\section{Introduction}
\label{sec:introduction}
\textcolor{black}{In recent years, numerous reports} have demonstrated the ability of \textcolor{black}{Huygens' and bianisotropic} metasurfaces (MS) to \textcolor{black}{perform} complex wavefront manipulation \textcolor{black}{at microwave frequencies} \cite{Pfeiffer2013, Asadchy2015}. These planar structures are composed of subwavelength \textcolor{black}{polarizable} elements (meta-atoms) \textcolor{black}{featuring electric, magnetic, and (possibly) magnetoelectric responses}. \textcolor{black}{Synthesis of such MSs usually includes a \emph{macroscopic} design stage, where the necessary homogenized MS constituent distribution is derived from the desired field transformation; and a \emph{microscopic} design stage,}
where this abstract distribution is discretized, and physical structures for the meta-atoms implementing the required local response are devised \cite{Epstein2016_2}. \textcolor{black}{The latter typically relies on time-consuming full-wave optimization; although alternative semianalytical transmission-line-model approaches have been proposed for the common cascaded impedance sheet formation \cite{Pfeiffer2014_3}, these techniques neglect interlayer near-field coupling effects, thus require a final optimization step in full-wave solvers \cite{Cole2018}.}

\begin{figure}[hbtp]
\center
\includegraphics[width=6cm]{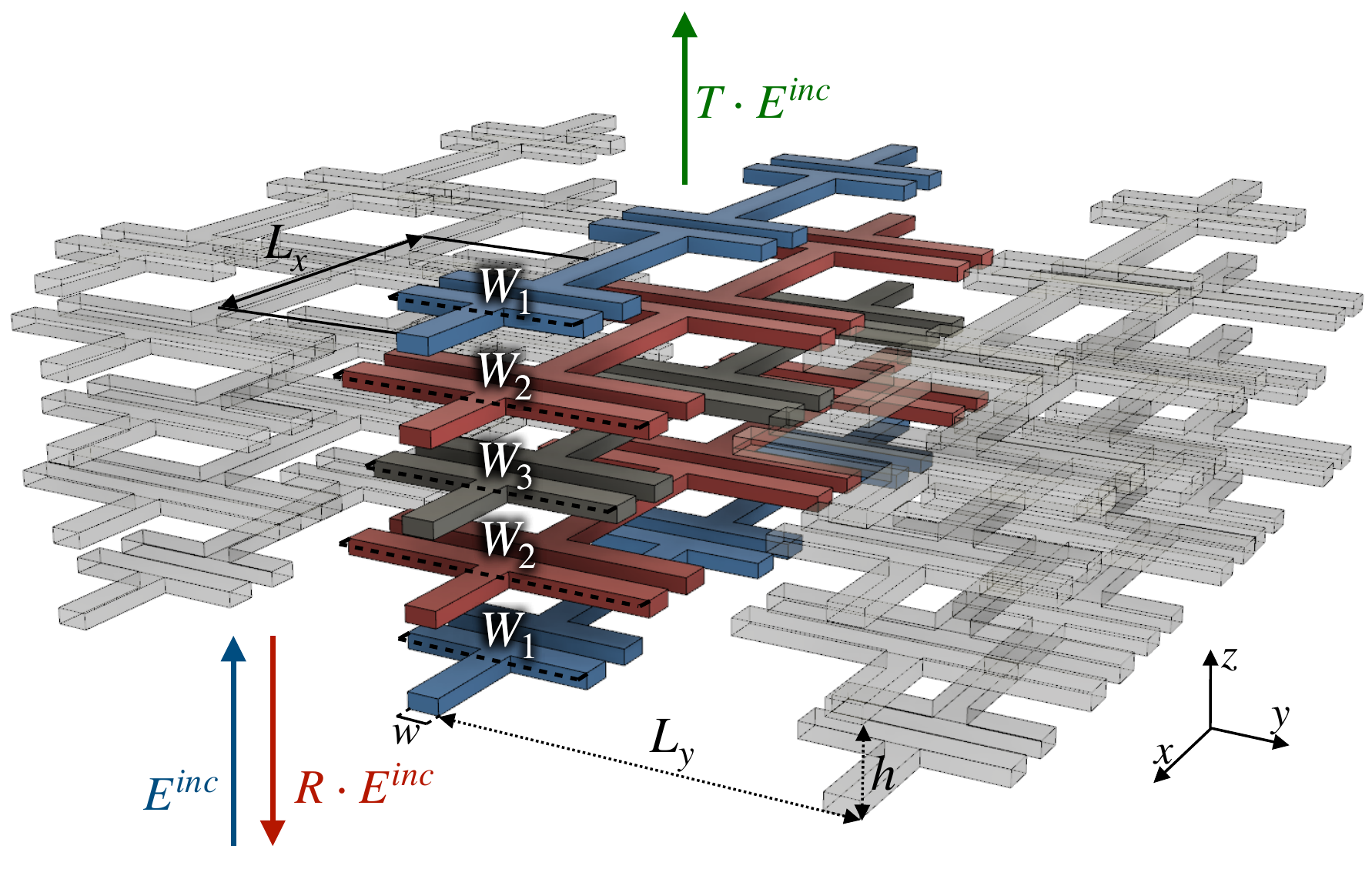}
\caption{\textcolor{black}{Multilayered meta-atom characterization configuration ($N=5$)}.}
\label{fig:physical_configuration}
\end{figure}

In this paper, we present a rigorous reliable microscopic design method to synthesize a multilayer \textcolor{black}{MS, composed of $N$ cascaded capacitively-loaded wires} (Fig. \ref{fig:physical_configuration}). 
\textcolor{black}{The method relies on the analytical model presented in \cite{Ikonen2007} and adjusted for metagratings \cite{Epstein2018}, allowing characterization of the meta-atom response in a way that properly accounts for the near-field phenomena. We utilize this method to semianalytically compose a look-up table (LUT) for 5-layer Huygens' meta-atoms, used to design a Huygens' MS (HMS) for anomalous refraction. Simulations in a commercial solver verify that, even for extremely close interlayer separation, highly-efficient refraction is obtained, without any full-wave optimization.} 

\section{Theory}
\label{sec:theory}
We consider \textcolor{black}{multilayered meta-atoms composed of $N$ cascaded wires, loaded with printed capacitors of width $W_n$, positioned at the planes $z=nh$ ($0\leq n<N$), and designated to be excited with transverse electric fields (${E_z=E_y=H_x=0}$). To associate a given physical structure with effective polarizabilities (thus compiling the aforementioned LUT), the $L_x\times L_y$ meta-atoms are repeated periodically along the $x$ and $y$ axes ($L_x,L_y\ll\lambda$), and their response to a normally incident plane wave $E_x^{\mathrm{inc}}\!\left(\vec{r}\right)\!=\!E_\mathrm{in}e^{-jkz}$ is evaluated (Fig. \ref{fig:physical_configuration}) \cite{Epstein2016_2}.} 


\textcolor{black}{When excited by $E_x^{\mathrm{inc}}$, currents will be induced on the wires at the $n$th layer; as $L_x\ll\lambda$, these can be effectively treated as electric line sources carrying current $I_n$ parallel to the $x$ axis, uniformly loaded by an impedance density of $\tilde{Z}_n$ (associated with the printed capacitor of width $W_n$) \cite{Ikonen2007, Radi2017, Epstein2018}. Consequently, the fields produced by the excited wires at $z=nh$ read} 
\begin{equation}\label{eq:onelayer}
E_{x,n}^{\mathrm{ls}}\!\left(\vec{r}\right)\!=\!-\frac{k\eta}{4}I_{n}\!\!\!\!\sum_{p=-\infty}^{\infty}\!\!\!\!H_{0}^{\left(2\right)}\!\!\left[k\sqrt{\left(y-pL_y\right)^{2}+\left(z-nh\right)^{2}}\right]\!\!\!
\end{equation}
\textcolor{black}{where the induced currents $I_n$ need to be computed to assess the scattered fields. This can be achieved by invoking Ohm's law on the wires; using the Poisson formula, one gets \cite{Ikonen2007,Epstein2018}}
\begin{eqnarray}\label{eq:current}
\tilde{Z}_{n}I_{n}\!\!\!&=&\!\!\!\!\!E_{\mathrm{in}}e^{-jknh}+j\frac{k\eta}{2{\pi}}I_{n}\log\left(\frac{\pi w}{2 L_y}\right) \nonumber\\ 
\!\!\!\!\!\!&-&\!\!\!\!\!\frac{\eta}{2L_y}\left(\sum_{l=0}^{N-1}I_{l}e^{-jk\left|n-l\right|h}\right)\nonumber\\ 
\!\!\!\!\!\!&-&\!\!\!\!\!jk\eta\sum_{m=1}^{\infty}\left(\frac{\sum_{l=0}^{N-1}I_{l}e^{-\alpha_{m}\left|n-l\right|h}}{L_y\alpha_{m}}-\frac{I_{n}}{2{\pi}m}\right)
\end{eqnarray}
\textcolor{black}{where $w$ is the wire width and $\alpha_m=\sqrt{(2\pi m/L_y)^2-k^2}$.} 

\textcolor{black}{For given meta-atom geometry, i.e. given load impedances $\tilde{Z}_n$, \eqref{eq:current} forms a set of linear equations, from which the currents $I_n$ can be resolved. Using these, the total scattered fields can be evaluated by summing the contributions \eqref{eq:onelayer} of the $N$ cascaded layers and the incident field. Finally, with the Poisson formula, the total fields can be expressed as a sum of Floquet-Bloch (FB) modes, from which the homogenized (zeroth-order mode) transmission and reflection coefficients associated with the given meta-atom can be readily retrieved as \cite{Ikonen2007,Epstein2016_2}}
\begin{equation}\label{eq:coef}
\begin{array}{l l}
\!\!T\!=\!1\!-\!\frac{\eta}{2L_y}\!\!\sum_{n=0}^{N-1}\!\!\!\frac{I_{n}}{E_{\mathrm{in}}}e^{jknh}, &
R\!=\!\frac{\eta}{2L_y}\!\!\sum_{n=0}^{N-1}\!\!\!\frac{I_{n}}{E_{\mathrm{in}}}e^{-jknh}
\end{array}
\end{equation}

\section{Results and Discussion}
\label{sec:results}
\textcolor{black}{To verify and demonstrate the prescribed formalism, we utilize it to design a HMS for anomalous refraction, redirecting a normally incident plane wave towards $\theta_\mathrm{out}=56^\circ$. The MS is composed of a \emph{symmetric} cascade of $N=5$ layers, featuring 3 degrees of freedom, $W_1, W_2, W_3$, with $L_x=L_y=\lambda/5$ and $h=\lambda/20$ (Fig. \ref{fig:physical_configuration}); the macro-periodicity required to obtain the desired refraction is $\Lambda=1.2\lambda$ (6 unit cells).}

\textcolor{black}{Our first goal is to compile the suitable LUT, associating all feasible triplets $\left(W_1, W_2, W_3\right)$ with their corresponding unit-cell transmission and reflection coefficients. To use \eqref{eq:onelayer}-\eqref{eq:coef} to this end, we first need to establish a relation between the capacitor width $W$ and the effective load impedance-per-unit-length $\tilde{Z}$. This is done by simulating a single-layer structure in ANSYS HFSS for a range of $W$ values, and extracting via \eqref{eq:current}-\eqref{eq:coef} with $N=1$ the \emph{complex} $\tilde{Z}\left(W\right)$ leading to the recorded scattering parameters. Equipped with the interpolated relation $\tilde{Z}\left(W\right)$, we sweep the design space $\left(W_1, W_2, W_3\right)$ and use MATLAB to calculate for each triplet the predicted $R$ and $T$ following the scheme in Section \ref{sec:theory}, forming the desired LUT.}

%

\begin{figure}[t]
\centering
\includegraphics[width=8cm]{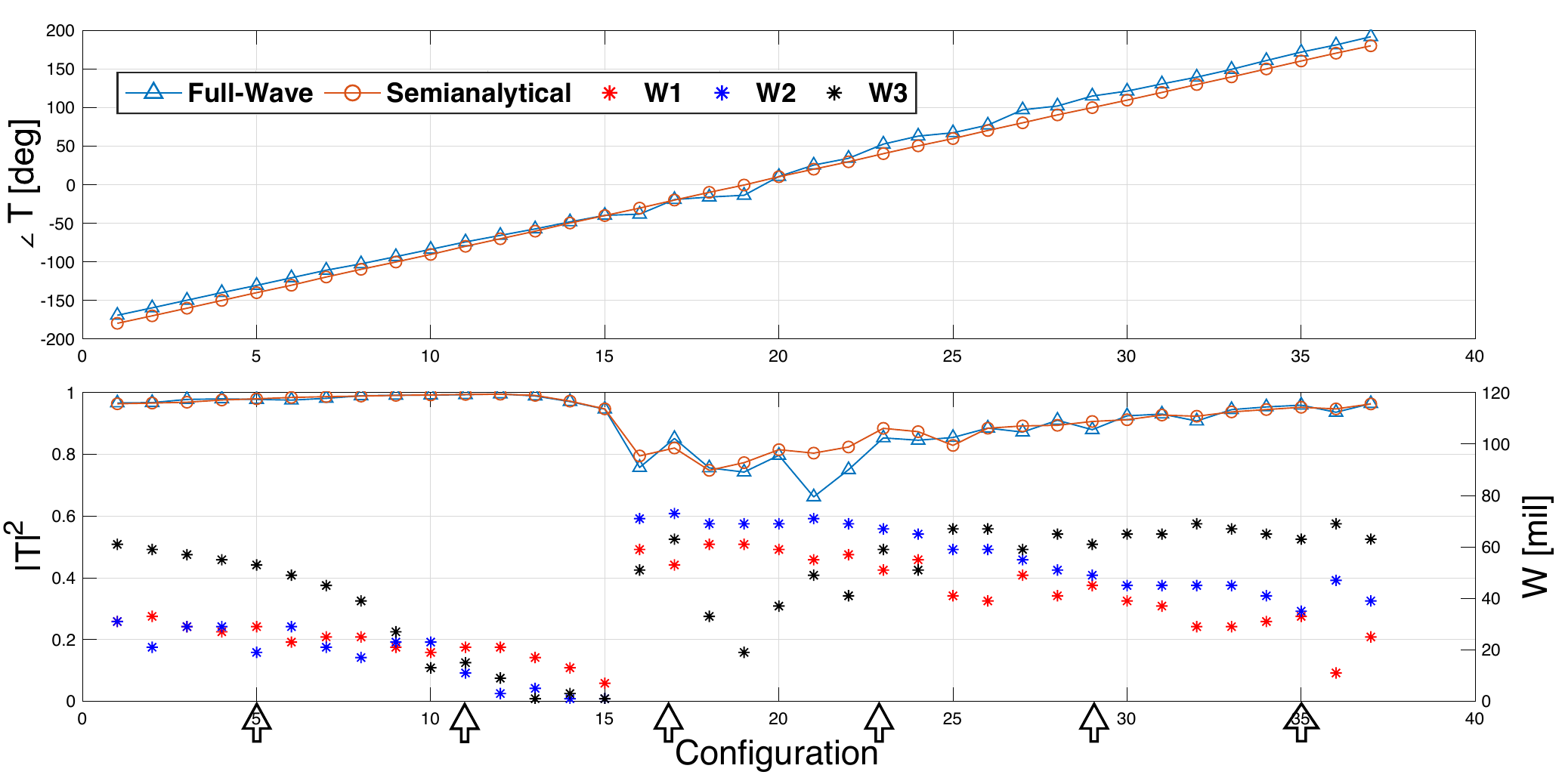} 
\caption{\textcolor{black}{Comparison between analytically predicted and full-wave simulated transmission (a) magnitude and (b) phase, for the capacitor widths $\left(W_1, W_2, W_3\right)$ denoted by * markers (Fig. \ref{fig:physical_configuration}).}}
\label{fig:LUT}
\end{figure}

\textcolor{black}{For refraction involving a normally incident plane wave, $0^\circ$-Huygens' meta-atoms can be used \cite{Epstein2016_2}, corresponding to unit cells exhibiting unity transmission magnitude and linear phase across the period. Hence, we use the LUT to identify those combinations $\left(W_1, W_2, W_3\right)$ yielding the maximal transmission magnitude for every possible phase shift. The best unit cell geometries as a function of $\angle T$, obtained without any full-wave optimization, are presented in Fig. \ref{fig:LUT} (* markers), along with the predicted (orange circles) transmission magnitude and phase. The blue triangles denote the simulated values extracted for each of these meta-atoms from HFSS match these predictions well, indicating the fidelity of our analytical model, clearly capable of incorporating realistic copper losses.}

\begin{figure}[hbtp]
 \centering
\includegraphics[width=4cm]{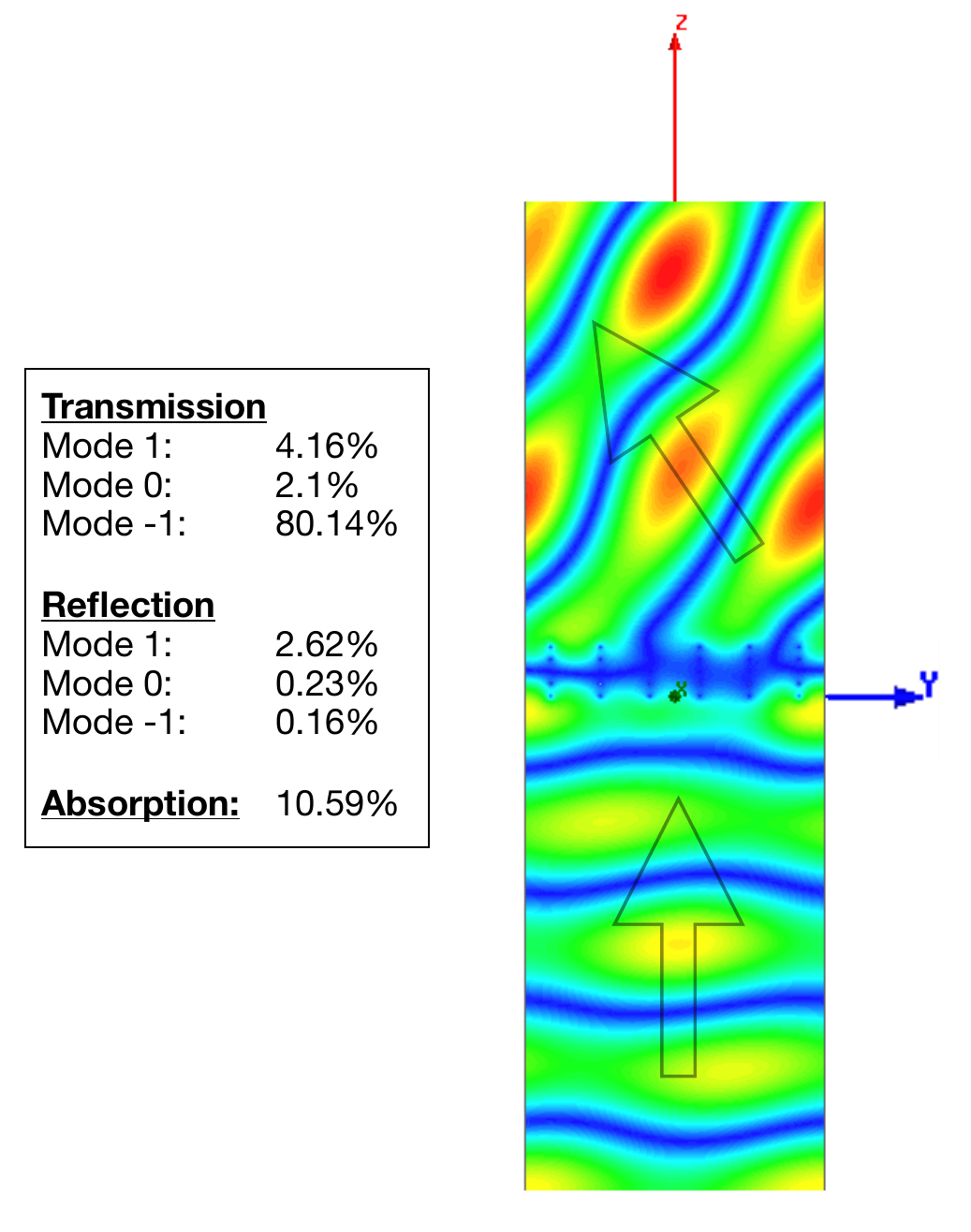} 
\caption{\textcolor{black}{Full-wave simulation results for the anomalous refraction HMS ($\theta_\mathrm{in}=0^\circ$, $\theta_\mathrm{out}=56^\circ$) semianlytically designed in Section \ref{sec:results}.}} 
\label{fig:Refractor}
\end{figure}

\textcolor{black}{Lastly, to form the desired MS, we sample Fig. \ref{fig:LUT} in $60\deg$ intervals, choosing these $6$ unit cells with the highest average $\left|T\right|$. This macro period, comprised of 30 copper capacitively-loaded wires with the $W_n$ prescribed by the LUT, was simulated in HFSS under periodic boundary conditions. The recorded electric field distribution and the coupling efficiencies to the various FB modes are presented in Fig. \ref{fig:Refractor}. These results clearly indicate that although no full-wave optimization was performed, our semianalytical design scheme yielded a highly efficient refracting HMS. In fact, if one excludes the inevitable conductor losses, it is found that about $90\%$ of the scattered power is coupled to the desired $-1$ mode in transmission, very close to the optimal coupling efficiency of $92\%$ \cite{Epstein2016_2}.}
 
\section{Conclusion}
\label{sec:conclusion}
To conclude, we presented \textcolor{black}{a rigorous semianalytical} scheme for \textcolor{black}{microscopic design of} multilayered MSs. \textcolor{black}{In contrast to previous techniques, the model properly accounts for interlayer near-field coupling, thus exhibiting high fidelity. Verified via commercial solvers, we expect this efficient methodology, avoiding full-wave optimization, to accelerate practical realizations of MSs for a variety of functionalities}.
%
%

\bibliographystyle{IEEEtran}
\bibliography{MicroscopicDesignMS}

\end{document}